\documentclass{mn2e}
\usepackage{epsfig}
\usepackage{amsmath}
\usepackage{longtable}

\def\d{\mbox{d}}

\setlongtables

\title[]{Multi-dimensional simulations of radiative transfer in
Type~Ia supernovae}
\author[S. A. Sim]{S. A. Sim\\
Max-Planck-Institut f\"{u}r Astrophysik, Karl-Schwarzschildstr. 1,
85748 Garching, Germany}

\date{7 November 2006}

\pagerange{\pageref{firstpage}--\pageref{lastpage}}
\pubyear{2006}
\begin{document}
\maketitle
\label{firstpage}

\begin{abstract}
A three-dimensional 
Monte Carlo code for modelling radiation transport in Type Ia
supernovae is described. In addition to tracking Monte Carlo quanta
to follow the emission, scattering and deposition of radiative energy,
a scheme involving volume-based Monte Carlo estimators is used to
allow properties of the emergent radiation field to be extracted for
specific viewing angles in a multi-dimensional structure. This 
eliminates the need to compute spectra or light curves by angular
binning of emergent quanta.
The code is applied to two test problems to illustrate consequences of 
multi-dimensional structure on the modelling of light curves. First,
elliptical models are used to quantify how large scale asphericity can
introduce angular dependence to light curves. Secondly, a model which 
incorporates complex structural inhomogeneity, as predicted by modern
explosion models, is used to investigate how such structure may affect
light curve properties.
\end{abstract}


\section{Introduction}

Despite decades of study, type Ia supernovae (SNIa) continue to be an 
active topic for astrophysical research. The accepted physical
mechanism for these
events, that they are the result of thermonuclear explosions of
degenerate material (Hoyle \& Fowler 1960) in  white
dwarf stars, is very well established but owing to the considerable
complexity of this process many aspects of these spectacular events
remain poorly understood.

Observationally, it is clear that all SNIa are not the same --
there is significant diversity in both brightness and decay
timescale (Phillips 1993). Aside from its direct relevance to the study of
SNIa themselves, understanding this diversity has important
implications to other branches of astrophysics -- in particular to
cosmology where inferences about the properties of distant SNIa play
an important role in probing the rate of expansion of the Universe.
Some correlations between supernova properties 
are already well established from observations of
nearby SNIa (e.g. Phillips 1993). However, it is becoming clear
that the observed diversity is not adequately
described by a single parameter (Benetti et al. 2004, 2005) 
and that unraveling this diversity
relies on a combination of careful observational study and sophisticated
theoretical modelling.

In recent years, there has been rapid development in the sophistication
of numerical modelling of the hydrodynamics of SNIa explosions. In
particular, while
in earlier work supernovae explosions were modelled using first
one-dimensional (1D) computer codes (e.g. Nomoto, Thielemann \& Yokoi 1984;
H\"{o}flich, Wheeler \& Thielemann 1998) and later two-dimensions
(e.g. M\"{u}ller \& Arnett 1986; Niemeyer, Hillebrandt, Woosley 1996), the most
up-to-date simulations are fully three-dimensional 
(3D, e.g. Reinecke, Hillebrandt \& Niemeyer 2002; Gamezo et al. 2003; R\"{o}pke
2005;
R\"{o}pke et al. 2006). These
multi-dimensional models are crucial for the understanding of
realistic flame propagation and hence nucleosynthesis in SNIa
explosions and have clearly demonstrated that the earliest 1D
models greatly underestimate the likely complexity of real SNIa. This
naturally raises the question of whether and to what extent this 
multi-dimensional complexity may be responsible for the observed diversity
of explosions.

To connect hydrodynamical explosion models and observations of real
SNIa light curves or spectra requires modelling of radiation
transport in the supernova. 
The majority of the light escaping from a supernova explosion at
around optical maximum
originates from energy deposited in the ejecta by the absorption and
Compton scattering of 
$\gamma$-rays emitted by radioactive isotopes. 
Given that all SNIa explosion models are
significantly optically thick at the earliest times ($\leq
1$~day), the emission, transport and deposition of radiation 
need only be followed for times after 
most of the complex dynamics have ceased and the ejecta is in
near-homologous expansion; this greatly simplifies the radiation
transport calculation and allows it to be decoupled from the
hydrodynamical simulation.

The level of sophistication of modern radiation transport computations
for supernovae 
has gradually developed to keep pace with the development of 
the hydrodynamical explosion models: for earlier applications
(e.g. Branch et al. 1981, Lucy 1987, Mazzali \& Lucy 1993) simple
one-dimensional calculations were sufficient but there is now growing 
interest in
fully three-dimensional simulations.
In particular, Lucy
(2005) has recently described and tested a sophisticated Monte Carlo
approach to 3D, time-dependent radiation transport. 
This method holds much promise for a wide
range of applications and is already being adopted and developed in
several contexts 
-- for example, Kasen, Thomas \& Nugent (2006) have developed a 
code based on that described by Lucy (2005) which incorporates a 
treatment of polarisation and lifts the
assumption of grey opacity used in the earlier code.
Independently, Maeda, Mazzali \& Nomoto (2006) have also adopted Lucy's 
techniques and utilized them in multi-dimensional simulations of light
curves for core collapse supernovae.

In this investigation, Monte Carlo calculations using methods similar
to 
those
of Lucy (2005) are used to investigate aspects of the influence of the
3D structure of SNIa on their observable properties. In Section~2,
the computer code used for the calculations is described. Although
similar to the codes described by Lucy (2005) and Kasen et
al. (2006), 
there are key
differences, specifically in the means by which the observable light
curves are extracted from the Monte Carlo simulation. This
code is then used to investigate the effect of departures from
spherical symmetry for two physically distinct cases.
First, in Section~3 toy models are used to investigate the
influence of large scale (low-mode) asymmetries on observable light
curves.
Secondly, a representation of a 
modern 3D explosion model is used to investigate the
implications of the complex, relatively small-scale inhomogeneity
predicted (Section~4).
The emphasis here is on identifying and interpreting differential
effects between 3D and 1D radiative transfer and thus several
simplifying assumptions are made in the treatment of the micro-physics.
Conclusions are drawn in Section~5.

\section{Method}

The calculations presented in this paper have been performed using a
Monte Carlo code which follows the propagation of radiative energy
in 3D as a function of time. This code is closely based on that
described by Lucy (2005) and thus only a brief summary of the
operation of the code is given here, with particular emphasis on the
the departures from the 
approach described by Lucy (2005).
The nomenclature used by Lucy for the Monte Carlo quanta (radioactive
pellets, $\gamma$-packets, $r$-packets) is adopted throughout.

\subsection{Conceptual summary of the code}

To obtain light curves for a pre-specified supernova
model, the code undertakes the following steps.
A computation domain is assigned which is large enough to 
encompass the physical extent of the model.
This is divided into a number of 
grid cells each of which is assigned an initial mass-density
based on the input supernova model.  
Pellets of radioactive material are placed in these cells, again in
accordance
with the chosen model.
The radiative decays of these pellets by emission of $\gamma$-rays and
the subsequent propagation and thermalisation of the $\gamma$-packets is
followed via a Monte Carlo calculation. A grey-opacity treatment is
adopted for the propagation of radiation ($r$-packets) in all other wave-bands.

During the Monte Carlo calculation, estimators are determined for
various properties of the radiation field in every grid cell (see
below). No information is retained regarding the variations of these
properties within individual grid cells.

After the Monte Carlo simulations are complete, the estimators are
used to determine observer frame emissivities. A formal solution of
the radiative transfer equation is then performed using these
emissivities to determine the emergent radiation field. The code
currently produces two classes of output: ``bolometric''
(ultraviolet-optical-infrared, {\sc uvoir})-light curves
which are obtained from the behaviour of the $r$-packets, and
$\gamma$-ray spectra and light curves which are derived from the 
$\gamma$-packets. For the applications described in Sections~3 and 4,
only the $r$-packet light curves are needed, however the analogous means by
which $\gamma$-ray properties are obtained are also described here for
future reference.

\subsection{Tracking of quanta and the grid}

Following Lucy (2005), a regular Cartesian grid which expands
with time is used. However, we allow the grid to expand continuously
rather than in discrete jumps at the end of a each time-step. Since the
supernova ejecta is assumed to be in homologous expansion, this adds
little complexity to computing the propagation of the Monte
Carlo quanta and removes the need to check whether they skip
across boundaries due to the modification of the grid at the start of
each time-step. In the limit of small time-steps, this modification
has no effect on the results of the calculation.

\subsection{Extraction of spectra and light-curves}

The code described by Lucy (2005) obtained light curves by directly
counting the number of quanta that escaped the computational domain 
during each time-step.
However, as discussed by Lucy (2005), 
that approach is not the most efficient with regard
to minimizing Monte Carlo noise.
Significantly higher quality spectra and light curves can be obtained
by using the paths of the Monte Carlo quanta to compute estimators 
for the emissivities in the model which can then be used, {\it post
hoc}, to obtain the intensity via a formal solution of the radiative transfer 
equation. Methods using such an approach have previously been used and shown to
be highly successful (see e.g. Lucy 1999).

For spherical models -- such as the test case used by Lucy (2005) --
the method of counting packets can always be used reliably, its 
inefficiency being countered by the use of a large number of
quanta. For complex 3D structures, however, this approach becomes 
unsatisfactory since the angular dependence of the escaping radiation
field can only be addressed by angular binning of the quanta. This has
further negative impact on the signal-to-noise of the computed spectra
and rapidly becomes prohibitive if more than a few 
angle-bins are to be considered.

In contrast, the use of emissivity estimators and a formal solution of
the radiative transfer equation allow the spectra and light curves to
be computed correctly for individual lines-of-sight to the supernova
without introducing extra Monte Carlo noise. Thus such methods are
strongly favoured for the study of any models which depart from
spherical symmetry. The implementation of these methods for 
the calculation of {\sc uvoir} light-curves and $\gamma$-ray spectra
are described in the next two sub-sections.

\subsection{{\sc uvoir} light curves}

The light-curve for {\sc uvoir} radiation as viewed by a distant
observer in direction {\boldmath $\hat{n}$} is given by:

\begin{equation}
L_{\mbox{\scriptsize obs}}(t_{\mbox{\scriptsize obs}}, \mbox{\boldmath $\hat{n}$}) = 4 \pi \int \!\!\! \int
I_{\infty}(\mbox{\boldmath
$d$}, t_{\mbox{\scriptsize obs}}, \mbox{\boldmath $\hat{n}$}) \d A
\; \; .
\end{equation}
Here, the integral is performed over the plane perpendicular
to  {\boldmath $\hat{n}$} and $I_{\infty}(\mbox{\boldmath $d$},t_{\mbox{\scriptsize obs}},\mbox{\boldmath $\hat{n}$})$
is the emergent intensity of a ray which is destined to reach the
observer at time $t_{\mbox{\scriptsize obs}}$. The intensity
depends on $\mbox{\boldmath $d$}$, the impact-vector of the 
ray given by $\mbox{\boldmath $d$} = \mbox{\boldmath $r \times
\hat{n}$}$ where {\boldmath $r$} is any position on the ray trajectory.
$L_{\mbox{\scriptsize obs}}(t_{\mbox{\scriptsize obs}}, \mbox{\boldmath {$\hat{n}$}})$ is the luminosity the observer
would imply were they to assume that the emission was isotropic. The true
(unobservable) luminosity of the supernovae, $L$, is given by

\begin{equation}
L(t_{\mbox{\scriptsize obs}}) =\frac{1}{4\pi} \int \!\!\! \int_{\mbox{all {\boldmath $\hat{n}$}}}
L_{\mbox{\scriptsize obs}}(t_{\mbox{\scriptsize obs}}, \mbox{\boldmath $\hat{n}$}) \d\Omega \; \; .
\end{equation}
This $L$ is the quantity obtained by directly summing all emerging
packets in a Monte Carlo simulation.

To evaluate equation~1 for a particular viewing direction 
({\boldmath $\hat{n}$}) at a particular time ($t_{\mbox{\scriptsize obs}}$),
a large sample of rays (typically $10^5$)
are chosen such that they cover the complete range of {\boldmath $d$}
required by the size of the supernova ejecta. These rays are launched
simultaneously from a plane perpendicular to ({\boldmath $\hat{n}$})
behind the supernovae having initially zero intensity. The time of launch is
chosen such that the ray with $\mbox{\boldmath $d$} = 0$ crosses the 
coordinate origin at time $t_{\mbox{\scriptsize obs}}$.
For each ray,
the emergent intensity is determined by numerically solving the
radiative transfer equation along its trajectory:

\begin{equation}
\frac{\d I(\mbox{\boldmath $r$}, t^{\prime}, \mbox{\boldmath $\hat{n}$})}{\d s} =
\eta(\mbox{\boldmath $r$}, t^{\prime}, 
\mbox{\boldmath $\hat{n}$}) - \kappa(\mbox{\boldmath $r$}, t^{\prime}, 
\mbox{\boldmath $\hat{n}$}) I(\mbox{\boldmath $d$}, t^{\prime}, \mbox{\boldmath
$\hat{n}$})
\end{equation}
where $\d s$ is the element of pathlength along the ray while
$\eta$ and $\kappa$ 
are the observer frame, direction-, time- and
position-dependent emissivity and opacity, respectively. 
Since the trajectory is a light ray, the position \mbox{\boldmath
$r$} and the time $t^{\prime}$ are related by $\d \mbox{\boldmath
$r$} / \d t^{\prime} = c \mbox{\boldmath $\hat{n}$}$.
The emergent intensity, $I_{\infty}(\mbox{\boldmath
$d$},t_{\mbox{\scriptsize obs}},\mbox{\boldmath $\hat{n}$})$, is the value of 
$I(\mbox{\boldmath $d$}, t^{\prime}, \mbox{\boldmath
$\hat{n}$})$ at the point where the ray trajectory finally leaves the
supernova ejecta.

In all the calculations presented here, 
the {\sc uvoir} opacity per unit density
in a grid cell 
is assumed to be constant and isotropic in the co-moving frame. Thus
the observer frame opacity, $\kappa(\mbox{\boldmath $r$}, t^{\prime}, 
\mbox{\boldmath $\hat{n}$})$ can be readily calculated.
In contrast, the emissivity 
$\eta(\mbox{\boldmath $r$}, t^{\prime}, \mbox{\boldmath $\hat{n}$})$
is not known {\it a priori} and is obtained 
from a Monte Carlo simulation.

In the current version of the code, there are two distinct {\sc uvoir}
emissivity source terms: one due to thermalisation of $\gamma$-ray packets
($\eta_{\gamma}$) and one
due to scattering of {\sc uvoir} photon packets ($\eta_{r}$).
Following Lucy (2005), estimators 
for these emissivities in the co-moving frame for a particular grid cell 
during a particular time step are obtained from the Monte Carlo energy
packet trajectories using:

\begin{equation}
\eta_{\gamma}^{\mbox{\scriptsize cmf}} = \frac{1}{4\pi V \Delta t}\sum_{\mbox{$\gamma$-paths}} \kappa_{\gamma}^{\mbox{\scriptsize cmf}}
\epsilon (1 - 2{\mbox{\boldmath
$v.\hat{n}$}}_{p}/c) \d s
\end{equation}
and

\begin{equation}
\eta_{r}^{\mbox{\scriptsize cmf}} = \frac{1}{4\pi V \Delta t}\sum_{\mbox{$r$-paths}} \kappa_{r}^{\mbox{\scriptsize cmf}}
\epsilon (1 - 2{\mbox{\boldmath
$v.\hat{n}$}}_{p}/c) \d s \; \; .
\end{equation}
In equation~4, the summation runs over all the trajectories of
$\gamma$-ray packets within the cell (which has volume $V$) during the 
time-step (which has duration $\Delta t$); $\d s$ is the
trajectory length and $\epsilon$ is the packet
energy determined in the observer frame. 
The packet is travelling in direction $\mbox{\boldmath
$\hat{n}$}_{p}$ and the velocity of the ejecta at the mid-point of the
trajectory is $\mbox{\boldmath
$v$}$.
The co-moving
frame $\gamma$-ray thermalisation opacity ($\kappa_{\gamma}^{\mbox{\scriptsize cmf}}$)
is frequency-dependent and includes contributions from both
Compton scattering and photoelectric absorption.
Note that the Compton term accounts only for the rate at which
$\gamma$-rays transfer energy to Compton electrons (which are assumed
to thermalise in situ).
In equation~5, the summation is over {\sc uvoir} packet
trajectories and $\kappa_{r}^{\mbox{\scriptsize cmf}}$
is obtained from the adopted {\sc uvoir} opacity coefficient.

It has been assumed here that the emissivity is isotropic in the
co-moving frame and that terms $O(v^2/c^2)$ and smaller can be neglected.
Note that although the co-moving frame emissivity
($\eta^{\mbox{\scriptsize cmf}}$) is isotropic, the
observer frame emissivity ($\eta$) is not, owing to the
Doppler terms in the transformation between frames.

The scheme described thus far is valid provided that the grid cells
are individually optically thin. In practice, this condition is
violated at early times when the ejecta is compact and dense. 
At such times, it becomes unacceptable to assign a uniform emissivity
to each grid cell. This can be overcome, however, by weighting each
contribution to the emissivities 
with a factor which 
accounts for the probability of energy absorbed and re-emitted during
the related trajectory escaping to infinity.
The weighting factor used here, which must be applied individually to
each contributing pathlength included in the sums in
equations 4 and 5 is

\begin{equation}
w(\tau(\mbox{\boldmath $\hat{n}$}), \delta\tau) = \exp(- \tau(\mbox{\boldmath $\hat{n}$}) -
\frac{1}{2}\delta\tau) \frac{\exp(\delta\tau) - 1}{\delta\tau}
\end{equation}
where 
$\delta\tau ={\mbox{\boldmath $\hat{n}.\hat{n}$}}_{p} \kappa_{r} \d s$
and
$\tau(\mbox{\boldmath $\hat{n}$})$ is the total $r$-packet optical
depth from the mid point of the trajectory $\d s$ to the edge of the
supernova in the direction {\boldmath $\hat{n}$}.
This form of the weighting factor is valid provided that {\it either}
the total optical depth across a grid cell is small {\it or} that the
contributing pathlengths ($\d s$) are all small compared to the
typical length scale on which the physical properties (e.g. mass
density) of the model vary. Therefore, in calculations which make use
of these weighted estimators, the pathlengths that energy packets can
travel in a single step are not permitted to exceed a predetermined
maximum step size, $\d s_{\mbox{\scriptsize max}}$. 
For each time step
in the calculation, $\d s_{\mbox{\scriptsize max}}$ is set to {\it
either}
one-tenth the width of a grid cell {\it or} the distance corresponding to 
$\delta\tau = 0.1$ in the densest grid cell; the larger of these two
values is chosen. Furthermore,
when the weighted emissivities are used, the opacity term in
equation~3 must be neglected since the probability of the emitted
energy
escaping has already been addressed.

This weighting overcomes the problem of having optically thick cells
at early times but it has two significant drawbacks.
First, the required computations of $\tau$ are time
demanding and lead to a substantial increase in code execution time. 
Secondly, since the weighting factor is angle dependent, separate sets
of estimators are required for each viewing angle that is to be
investigated (in contrast, the unweighted estimators are independent
of {\boldmath $\hat{n}$}). 
Therefore, this method should only be used for computing the light-curve
at early times -- when the density in the grid cells is high -- and be
replaced with the unweighted scheme once the 
opacity of individual cells has dropped sufficiently.

\subsection{Test calculations of the {\sc uvoir} light curve}

To test the implementation of the method described above, the code has
been applied to the test model used by Lucy (2005). This is a
spherical model, based on that used by Pinto \& Eastman (2000a),
having
total mass $M = 1.39 M_{\odot}$, $^{56}$Ni mass
$M_{^{56}\mbox{\scriptsize Ni}} = 0.625 M_{\odot}$ and a maximum velocity of
$10^4$~km~s$^{-1}$. The distribution of Ni is centrally peaked. 
Following Lucy (2005),
the
grey absorption cross-section of 0.1~cm$^{2}$~g$^{-1}$ is adopted for
{\sc uvoir} radiation, and the photoelectric absorption coefficient is
taken from Ambwani \& Sutherland (1988), adopting a mean value of $Z=14$. 
A 100$^3$ Cartesian grid is adopted for this test and $2.5 \times
10^6$ Monte Carlo packets were used in the calculations.

\begin{figure}
\epsfig{file=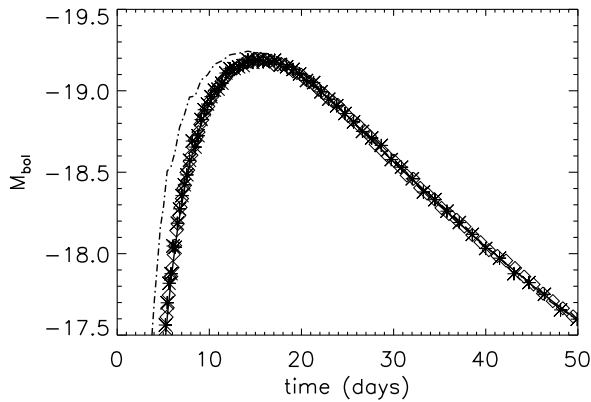, height=6cm}
\caption{
Comparison of {\sc uvoir}-light curves computed for the test model 
described in Section~2.5.
The solid line shows the light curve computed by Lucy (2005).
The open diamond symbols indicate the light curve obtained by counting 
emergent packets with the code developed here. The broken line is the
{\sc uvoir}-light curve computed using the estimators defined by
equations 4 and 5. The star symbols show the light curve obtained
using the weighted estimators (Section 2.4, equation~6).
}
\end{figure}

The light curve obtained by Lucy (2005) for this model is shown in
Figure 1 (solid line). This agrees very well with the light curve
computed here by directly counting the number of Monte Carlo packets
escaping this model (diamond symbols in Figure~1). It also matches
well with the light-curve computed for a particular viewing angle
using the weighted estimators described in Section 2.4
(the particular viewing direction was chosen randomly here).
However, when the estimators are not weighted (i.e. if equations 4
and 5 are used directly), the computed light curve is significantly 
overestimated at early times ($< 20$ days, in this case). 
As discussed in Section 2.4, this is because the grid cells of the model
are optically thick at early times; the maximum optical depth across
one grid cell is plotted, as a function of time, in Figure~2. This
shows that significant errors arise if the unweighted estimators are
used when this optical depth is
$\tau_{\mbox{\scriptsize cell}} > 1$.

\begin{figure}
\epsfig{file=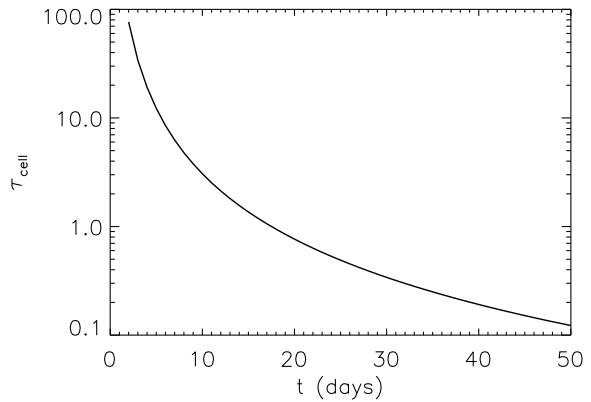, height=6cm}
\caption{
The maximum {\sc uvoir} optical depth across a grid cell as a function
of time in the test model described in Section~2.5.
}
\end{figure}

\subsection{$\gamma$-ray spectra and light curves}

The $\gamma$-ray spectrum is obtained following the same principles as
that used for the {\sc uvoir} light curve described above. The
important difference being that the treatment of $\gamma$-rays is
fully frequency dependent.

To deal with the frequency dependence, a grid of frequency points is
used to divide the spectrum into small frequency intervals. 
One point in this grid is set to the rest frequency of each of the 
radioactive $\gamma$-ray emission lines in the range of interest and
the remaining points are spaced logarithmically between. To compute
the spectrum, the same scheme of solving the radiative transfer
equation along a set of rays trajectories through the model 
is used. Here, however, 
the radiative transfer along each trajectory is computed multiple
times, once for each
frequency point in the frequency grid, thereby determining
the emergent radiation field as a function of both frequency and time.

As for the grey-computations described above, the opacity term in the
radiative transfer equation is known (the sum of Compton and
photoelectric terms).
There are again two emissivity terms which need to be considered. 
The first, direct emission of $\gamma$-rays by radioactive decay
can also be expressed analytically for every grid cell in terms of the
half-lives of the radioactive isotopes and their initial
concentrations
in the cell.

The second emissivity term is due to Compton down-scattering. The
treatment of this term requires that both the angular- and
frequency-dependence of the Compton process be considered. It is
determined via a set Monte Carlo estimators (one per frequency interval per
grid cell per time step); the estimator for the frequency interval $i$
in the
grid cell $j$ for timestep $k$ is given (in the observer frame) by

\begin{equation}	
\eta_{ijk} = \frac{n_{e}}{V_{j} \Delta\nu_{i} \Delta t_{k}}
	\sum_{\mbox{$\gamma$-paths}}
	\epsilon
	\left({\frac{\d \sigma}{\d \Omega}}\right)_{\mbox{\scriptsize
	obs}}
	\d s
\end{equation}	
where the sum runs over all $\gamma$-ray trajectories which lie
in cell $j$ (having volume $V_{j}$) and have frequency appropriate for
scattering into the frequency
interval $i$ (of width $\Delta\nu$) during the timestep $k$
(which has duration $\Delta t_{k}$).
$n_{e}$ is the number density of target electrons in the cell,
$\epsilon$ is the observer-frame energy of the
$\gamma$-ray packet and 
$\left({{\d \sigma}/{\d \Omega}}\right)_{\mbox{\scriptsize obs}}$
is the Compton differential cross-section for scattering into the
direction of the line-of-sight ({\boldmath $\hat{n}$}) in the observer
frame. This cross-section depends upon the angle between the
trajectory and {\boldmath $\hat{n}$} and is determined by
applying the Klein-Nishina formula for the cross-section in the
co-moving frame.

\section{Application 1: An ellipsoidal model}

In this section, two toy models of elliptical supernovae are used to
investigate possible observational consequences of large scale asphericity 
in supernova explosions. Such an investigation is motivated by
observational evidence for global asphericity in SNIa obtained via
polarimetry (see e.g. Howell et al. 2001; Wang et al. 2003).
The origin of this asphericity is not well known: mostly likely it is
determined by the details of the explosion process itself but may have
its roots in the properties of a rapidly rotating progenitor. 
Here, however, the objective is not to gain insight to the physical
origin of such a geometry but rather to study how it might effect both the
amplitude and shape of observed light curves in comparison with
spherical explosions.

Earlier calculations of radiation transport in elliptical supernovae
have been discussed by H\"{o}flich (1991). In that work, a
multi-dimensional Monte Carlo code was also used. However, the
treatment of energy packet generation and emission was simplified via
the use of a parameterised photosphere in contrast to the full
treatment of $\gamma$-transport and deposition employed here.

\subsection{The model}

A simple elliptical model has been
constructed, closely related to the spherical model used as a test
case in Section~2. The adopted model has the same total mass and same 
$^{56}$Ni mass as the spherical model. It is also assumed to be in
homologous expansion and to have uniform density. However, the maximum
velocity (and hence spatial extent) is taken to be smaller
in the $z$-direction than in the $x$- and $y$-directions (symmetry is
still assumed under rotation about the $z$-axis). Such a model is
intended as a simple description for cases in which either the
explosion mechanism or the properties of the progenitor lead to a
large-scale (low angular mode) departure from sphericity.

Two particular realisations of the model will be considered here.
For both 
the maximum velocity in the $x$- and $y$-directions was kept at 
10$^9$~cm~s$^{-1}$.
The models differ in the chosen maximum velocity in the $z$-direction:
for one, this velocity was fixed
at $5 \times 10^{8}$~cm~s$^{-1}$, thereby giving an
axis ratio of 2:1; while for the other $8 \times 10^{8}$~cm~s$^{-1}$
was adopted to give an axis ratio of 5:4.
The axis ratio of 2:1 may be regarded as extreme -- it is comparable
to the axis ratio that might be present in a very rapidly rotating
progenitor (see e.g. Yoon \& Langer 2005) but there is little evidence 
to suggest that this would be preserved during an explosion. 
As an extreme case, however, this model is useful for providing a
clear indication of the sense in which asphericity can affect the
light curve.
The
second ratio adopted (5:4) is comparable to the $\approx 20$ per cent
asphericity found by Howell et al. (2001) for SN1999by and is
therefore more likely to be indicative of the propertied of real SNIa 
explosions.

As discussed in Section~1, the interest here is in probing the effect
of a 3D treatment of the radiative transfer compared to spherically
symmetric 1D calculations and so the simplifying assumption of a
constant, grey-{\sc uvoir} opacity is retained.
For all the calculations discussed in the section, this grey-{\sc
uvoir} absorption cross-section remains fixed at 
$\sigma = 0.1$~cm$^{2}$g$^{-1}$. 

\subsection{{\sc uvoir} light curves}

Light-curves have been computed for observers viewing the ellipsoidal 
supernovae from infinity along both the major- and minor-axes. These
were computed using the weighted-estimators method described in
Section~2.4.
The two light curves 
for the model with axis ratio 2:1 are plotted in Figure~3, along with the
angle-averaged 
light curve for the same model (i.e. the light curve obtained from the
arithmetic mean of the light curves seen by a large number of
observers from random viewing angles).

\begin{figure}
\epsfig{file=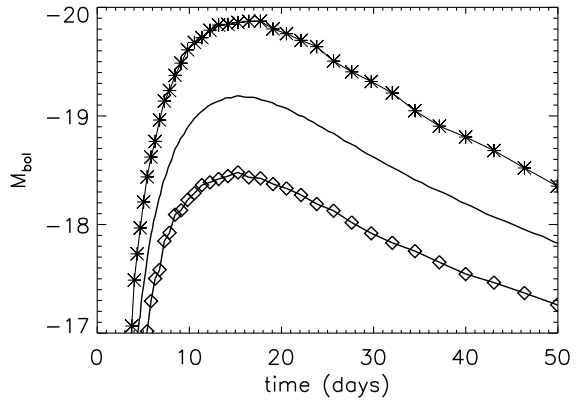, height=6cm}
\caption{
Computed light curves for the elliptical model with axis ratio 2:1. 
The solid line shows the
angle-averaged light curve, the diamonds show the results for viewing
down the long axis and the stars for viewing down the short axis.
}
\end{figure}

When viewed down the short-axis, the light curve is considerably
brighter than as observed down the long-axis. Around maximum light, 
the difference in brightness is approximately a factor of 3.5.
At later times, as the supernova becomes less optically thick, the
difference becomes smaller -- approximately
a factor of 2.5 at around 50 days. 
The light curve peaks slightly earlier when viewed down the long axis
(at $\sim 15.3$ days compared to $\sim 16.4$ days if viewed down the
short axis). Also, the light curve decays more slowly if viewed down
the long axis; this is characterised by the bolometric
$\Delta M_{15}$-values\footnote{Commonly used 
in analyses of light curves following
Phillips (1993), $\Delta M_{15}$ is the change in magnitude between
maximum light and 15 days after maximum light.}
of 0.55 (long-axis) and 0.64 (short-axis). 

Qualitatively similar, but quantitatively smaller differences are seen
in the light curves computed for different viewing angles using the
model with the axis ratio 5:4. For this model, representative light
curves
are show in Figure~4.

\begin{figure}
\epsfig{file=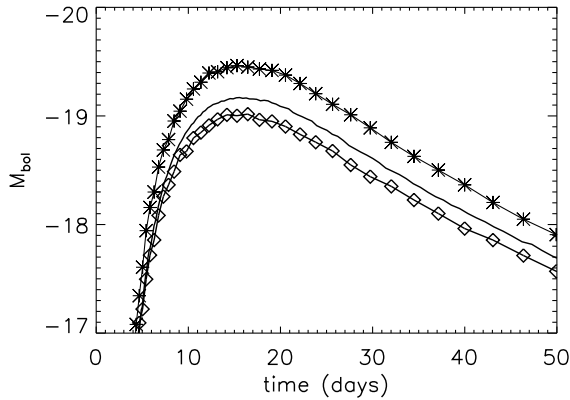, height=6cm}
\caption{
Computed light curves for the elliptical model with axis ratio 5:4. 
The solid line shows the
angle-averaged light curve, the diamonds show the results for viewing
down the long axis and the stars for viewing down the short axis.
}
\end{figure}

The scale of the angular variation is determined by the interplay of 
several effects. Consider viewing an
opaque ellipsoid with axis ratio 2:1 and uniform surface brightness as 
characterised by uniform surface temperature. One would expect to
find the flux to be twice as large when viewing along the short axis
compared to the long axis, simply due to the increase in apparent
surface area. The computed ratio exceeds this for two closely related
reasons. First, the choice of uniform density in the models means
that the column density to any particular 2:1 ellipsoidal surface
within the model is always less when viewing down the short axis than
the long. 
Thus the optical depth to the surface is smallest along the short axis
making the light curve brighter when viewed from this direction.

Secondly, contours of constant radiation density 
(or equivalently emissivity) do not exactly follow 
the 2:1 ellipsoidal geometry of the model; in the outer regions
there can be significant departures, always in the sense
that the radiation energy density is highest at the points of
intersection by the short axis. 
This is illustrated in Figure~5 where the variation of the $r$-packet 
emissivity ($\eta_{r}^{\mbox{\scriptsize cmf}}$) is shown along the
ellipsoidal axes.
A higher radiation energy density on the short axis makes the light
curve brighter when viewed down that axis and so enhances the
variation in brightness with viewing angle. 
The origin of this effect also lies 
in the lower column densities along
the short axis. 
At late times, a quasi-static description of
the radiation field becomes valid since packets are not trapped 
in the model for a significant number of timesteps. Under such
circumstances, the lower opacity along the short axis means that
photons preferentially diffuse in that direction, making the energy
density on an 2:1 ellipsoidal surface peak at the points of
intersection with the short axis.
At earlier times, this effect is enhanced by the time-dependent nature of the
calculations; 
fewer packets manage to reach outer ellipsoidal 
shells than would be predicted in a quasi-static description because 
they have had
insufficient time to diffuse far enough. The difference in diffusion time
means that this affects the energy
density along the long axis more significantly than along the short
axis and thus acts to enhance the effect expected from the
quasi-static case. 
As time passes, the shapes of the contours of radiation energy density 
evolve slightly and the outward decline becomes somewhat 
less steep (this can be seen
by comparing the $t \sim 12$~days and $t \sim 40$~days results in
Figure~5). However, throughout the time range considered here, a 
departure from 2:1 ellipticity remains.

\begin{figure}
\epsfig{file=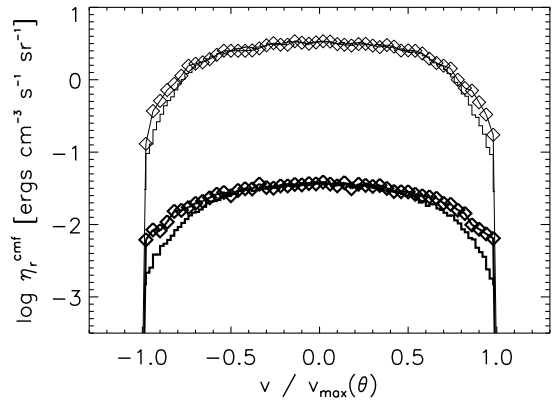, height=6cm}
\caption{
Variation of the $r$-packet emissivity, $\eta_{r}^{\mbox{\scriptsize
cmf}}$, in the ellipsoidal model with axis ration 2:1. 
$\eta_{r}^{\mbox{\scriptsize cmf}}$ is plotted along the short axis 
(connected-diamond symbols) and the long axis (histogram). 
These values were obtained from the Monte
Carlo estimators (equation~5) in the grid cells lying closest to the
axes. 
For ease of comparison, they are plotted against velocity in units of
the maximum speed ($v_{\mbox{\scriptsize max}}$) along the relevant
axis; thus if $\eta_{r}^{\mbox{\scriptsize
cmf}}$ were to follow the 2:1 ellipsoidal geometry of the model, the
diamond symbols and histogram would lie on the same curve.
Results are shown for two different times: the lightly drawn lines are 
for $t \sim 12$~days while the heavy lines are for $t \sim
40$~days. The $t \sim 40$~day values have been offset upwards by
+2~dex.
}
\end{figure}

Since both the effects described above are the result of angular
variations
in the optical depth to 2:1 ellipsoidal surfaces, both persist
while the ejecta remains optically thick. They slowly
decrease in strength during the decay phase as expansion causes the
optical depths to drop; at very late times the 
entire ejecta will become optically
thin to {\sc uvoir} radiation such that the light curve will become independent
of viewing angle. For the model adopted here, however, this nebular
phase will not begin until several hundred days later than the times
considered here.

The viewing-angle dependency of the light curves computed from these
simple models may have some interesting ramifications for
understanding the observed properties of SNIa light curves.
As pointed out by Wang et al. (2003) in the context of SN2001el,
directional dependence of the luminosity as predicted for elliptical
models of supernovae would lead to dispersion in the observed peak
magnitudes (based on the earlier work by H\"{o}flich 1991 and their
implied asphericity of $\approx 10$~per cent for SN2001el, they
speculate that this dispersion would be around 0.1~mag).
The results obtained here support this argument and indicate that if
the degree of asphericity were larger in some cases 
(e.g. SN 1999by; Howell et al. 2001) the spread in the peak
magnitude could be greater, $\approx 0.4$~mag.

Furthermore, the full light curves computed here allow this
dispersion relative to the known relationship between light curve
shape and peak luminosity to be examined.
This trend, the so-called ``Phillips relation'' following Phillips
(1993), expresses the negative correlation between
peak brightness and the $\Delta M_{15}$-parameter measured from
observed light curves of SNIa.
For both the models considered here,
the variation of $\Delta M_{15}$ with viewing angle is in the
opposite sense to the standard relationship. 
This is illustrated in Figure~6 where the six light
curves shown in Figures~3 and 4 are represented as points in the 
$\Delta M_{15}$-$M_{\mbox{\scriptsize peak}}$ plane. The gradient of
the standard
Phillips relation (describing the mean observed relationship between
$\Delta M_{15}$ and $M_{\mbox{\scriptsize peak}}$) in the B-band 
is plotted for comparison in the Figure.

This effect 
would lead to a detectable
scatter about the Phillips relation and thus may have a major role to play 
in understanding the diversity of supernova observations: the results
plotted in Figure~6 would suggest that if SNIa explosions were moderately
elliptical (such that the model with axis ration of 5:4 were
approximately applicable), viewing angle effects could explain
a scatter of several tenths of a magnitude about the mean relationship.
Significant 
caution must be applied in interpreting this result since the
grey-treatment adopted here does not allow band-limited light curves
to be studied for direct comparison with observations -- 
quantitative differences may occur if the frequency-dependence of the
opacity were taken into account.
Furthermore, the models used here have predicted an angular
variation of the radiation energy density and thus, by implication,
the temperature of the ejecta. Such a variation further contradicts
the used of a uniform opacity and highlights the need for the
consideration of more detailed micro-physics.
Also, the particularly simple model chosen (uniform density with
centrally concentrated $^{56}$Ni and time-independent mean opacity) 
produces significantly smaller
absolute values of $\Delta M_{15}$ than are typically observed -- thus
further work using more realistic models of aspherical supernovae
are needed.

\begin{figure}
\epsfig{file=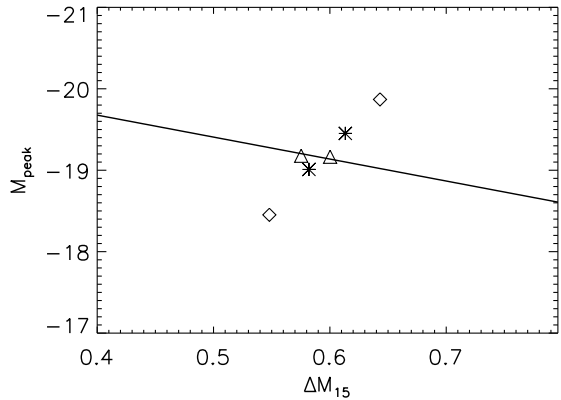, height=6cm}
\caption{
The figure shows the peak bolometric magnitude $M_{\mbox{\scriptsize
peak}}$ versus $\Delta M_{15}$ for the light curves computed from
the
elliptical SNIa models. The two points marked with diamonds indicate
the
light curves obtained for viewing the model with axis ratio 2:1
along its short and long axes. The stars indicate the light curves
corresponding to the same two viewing angles for the model with axis
ration 5:4. The triangles are for the angle-averaged light curves of
the two models. The solid line indicates the gradient of the observed 
relationship between $\Delta M_{15}$ and the peak magnitude {\it as
measured in the B-band} (Phillips 1993). The normalisation of the
observed relationship has been adjusted to approximately match the 
computed angle-averaged values.
}
\end{figure}

\section{Application 2: An inhomogeneous model}

Modern SNIa explosion models predict complex, three-dimensional
sub-structure (e.g. Reinecke et al. 2002; Gamezo et al. 2003; R\"{o}pke
2005;
R\"{o}pke et al. 2006) within the explosion. In contrast, most
models used to compute light curves for comparison with observation
have a smooth, one-dimensional density-/composition-profile. 

In this section, the effect of the predicted inhomogeneity on
model light curves will be investigated. In principle, there are two
classes of effect which are of interest here: first, inhomogeneity can
change the viewing-angle averaged light curve compared to that of a
spherically symmetric model.
Secondly, inhomogeneity could also lead to angular
dependence of the light curve -- the degree of angular dependence may
be reasonably expected to depend on a combination of the dynamic range
of the inhomogeneity in density and the physical length scale of the
variations. The length scale is relevant since if only small scale
inhomogeneity is present it will tend to be averaged out when
integrating over solid angle and so not introduce significant angular
variation of the light curve.

Given that the code used here adopts several simplifying
assumption -- most importantly, perhaps, that of a 
grey absorption coefficient for {\sc uvoir} radiation -- the emphasis 
here is on understanding
and assessing the differential effect of introducing 3D structure by
comparing with an equivalent 1D model using a fixed set of
well-understood approximations. Such a calculation is an important
first step in understanding the role of 3D structure and is a useful
starting point for further work where the micro-physics is
improved (see Section 5 for further discussion).

\subsection{The model}

The model used for this test calculation is based on the
3D explosion model computed by R\"{o}pke (2005). For this
model, R\"{o}pke (2005) followed the 
hydrodynamics of an exploding white dwarf star (with total mass
1.4~$M_{\odot}$) for 10 seconds on a 
schematic grid of 512$^{3}$ Cartesian cells. 
Only one spatial octant was 
simulated and symmetry under reflection was assumed to describe the
remaining octants (thus numerically only 256$^{3}$ grid cells were used). 
The distribution of mass-density and mass-fraction of iron-group
elements in each grid cell at the end of their
simulation was made available for this work. 
Light curves have previously been simulated from 1D representations of
this model by Blinnikov et al. (2006).

In order to make the
calculations here tractable -- in terms of both computer memory required and
photon statistics in each grid cell -- the model adopted here uses
only 170$^3$ grid cells. It was obtained from the 512$^3$ grid by
first removing the outermost cell from both ends of the grid
in each of the three Cartesian directions resulting in a 510$^3$
grid. The mass in the cells removed in this process was
negligible. The resolution was then reduced by a factor of 3 by
subdividing the 510$^3$ grid into 3$^3$ blocks and replacing each
block with a single cell whose density was equal to the mean density
of the original 27 cells. 

Given the grey treatment of {\sc uvoir} radiation currently adopted in
the code, it is not necessary to specify the detailed composition of
the material in each grid cell. However, it is necessary to specify
the initial distribution of $^{56}$Ni which provides the source of
radiative energy. The hydrodynamics code used by  R\"{o}pke (2005) provides as estimate of the fractional mass of iron-group
elements in each grid cell but does not give a reliable estimate of
the breakdown of this material into specific isotopes and elements.
The $^{56}$Ni mass-fractions used here were obtained by adopting a
constant ratio for the mass of $^{56}$Ni to the total mass of
iron-group elements in all grid cells. This ratio was fixed to yield a
total $^{56}$Ni-mass of 0.28~$M_{\odot}$, as derived by Kozma et
al. (2005) for this model; note that the nucleosynthesis calculations
described by Kozma et al. (2005) do not produce compositional
information in sufficient detail to reconstruct the full
3D-distribution of  $^{56}$Ni-mass owing to the modest (27$^3$) number
of tracer particles they used in comparison to the number of
grid cells in the 3D model.

In order
that the differential effect of the 3D structure could be assessed, a
1D comparison model was made by averaging the 3D model over
spherical shells, taking care to conserve the radial distributions of 
total mass and $^{56}$Ni-mass. In order to illustrate the degree of
inhomogeneity in the 3D model, Figure~7 shows the velocity 
distribution of density in the spherically averaged model with points
indicating the actual densities of individual grid cells in the 3D
model. This shows that across most of the velocity range of the model,
there is a spread in density of at least a factor of three
between different cells with similar velocities. 
This is comparable to the dynamic range of density for a
given velocity
in the original hydrodynamical model; thus one may be confident that the
model used here contains a fairly reliable representation 
of the inhomogeneity
implied by the hydrodynamics.

\subsection{Treatment of opacity}

\begin{figure*}
\epsfig{file=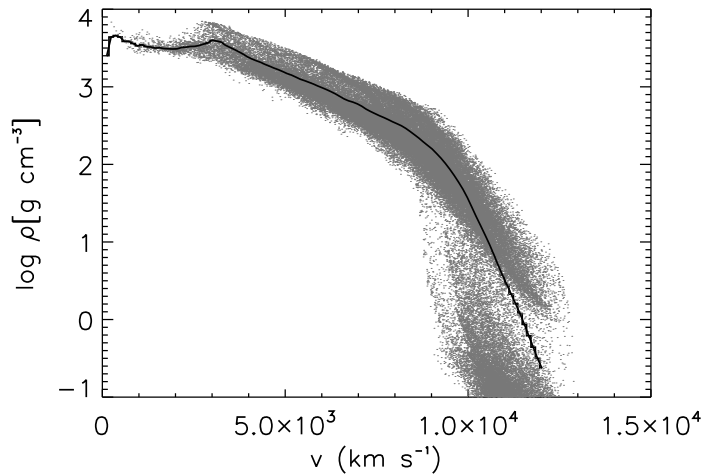, width=10cm}
\caption{
The solid line shows the density distribution with velocity of the
spherically averaged 1D model. The grey points each indicate the
density and mid-point velocity of a grid cell in the 3D model. 
All densities are shows for time $t = 10$~sec. 
}
\end{figure*}

We continue to adopt a grey {\sc uvoir}-absorption 
cross-section. However, to correctly evaluate the influence of
3D structure on the light curve, it is necessary to consider that
compositional inhomogeneity would cause
the cross-section per gram to
be a function of position. 
It goes beyond the scope of this paper to undertake full calculations
of the composition dependence of the opacity; instead, a simple
one-parameter description of the opacity is adopted, following 
Mazzali \& Podsiadolwski (2006). They consider the opacity to be
determined by the mass fraction of iron-group elements on the
assumption that the opacity per gram is a factor of ten higher for the
iron-group than for lighter elements. Adopting the same assumption, the
{\sc uvoir}-absorption cross-section per gram used in this section is given by

\begin{equation}
\sigma = N (0.9 X_{\mbox{\scriptsize Fe-grp}} + 0.1) \; \; 
[\mbox{cm$^{2}$ g$^{-1}$}] 
\end{equation}
where $X_{\mbox{\scriptsize Fe-grp}}$ is the mass fraction of iron-group
elements, which varies from cell to cell, and the normalisation factor
$N$ is chosen such that the mean value of $\sigma$ in the ejecta 
is fixed to $< \sigma > = 0.1$~cm$^{2}$ g$^{-1}$. 
Although crude, this parameterisation captures the essential physics
that the heavy elements dominate the mean opacity 
and only requires the compositional information which is directly
available from the explosion model (namely, the total iron-group mass).
Note that, for ease
of comparison with the constant-$\sigma$ calculations used in the
earlier sections of this paper, the
time-dependence imposed on $\sigma$ by Mazzali \& Podsiadolwski (2006)
is not included here.

In view of the simplifications used -- in terms of both numerical
resolution and particularly the very simple treatment of opacity --
the results obtained here should not be regarded as a definitive
prediction of the radiation properties of the R\"{o}pke et al. (2005) 
explosion model.
Rather, the emphasis is on using a model which has a reliable
representation of the degree of inhomogeneity in a real explosion
model to understand the role played by the complex structure in the
radiation transport.

\subsection{{\sc uvoir} light curves}

The light curve obtained with the 3D model described
above is shown as the solid histogram in Figure~8; this is the light
curve averaged over viewing angles. 
At peak, this light curve is dimmer than
those plotted in Figures~1, 3 and 4; a consequence of the lower
$^{56}$Ni-mass in the present model. The light curve peaks earlier (at
$t \approx 13$~days) as a result of differences in the 
distribution
of $^{56}$Ni with velocity: in the model used here, the
$^{56}$Ni-distribution is 
less centrally
concentrated than that adopted in Sections~2.5 and 3.

To establish the influence of the inhomogeneity on the light curve,
the 1D model obtained by spherically averaging the 3D model (see above)
was also used to compute a light curve.
This light curve is shown by the dashed curve in Figure~8.

From the Figure, it is apparent that the inhomogeneity causes the
light curve to be a little brighter at early times and very slightly
fainter during the decay phase. This occurs because in the
complex 3D medium a fraction of the $r$-packets find trajectories that
preferentially pass through lower density cells and therefore
encounter lower opacities than is possible in a spherically symmetric
model with the same total mass. This allows some $r$-packets to escape more
quickly making the light curve brighter during the pre-maximum
rise phase. It also means that slightly fewer packets remain trapped
in the ejecta which leads to the marginal dimming in the post-maximum
phase. The scale of the effect in this model is rather modest: at most it
amounts to a difference of $\approx 0.1$~mag and this occurs only at
very early times, $\approx 10$~days before maximum light. At around
maximum light, the effect is limited to a change in magnitude of
around  $\Delta M \approx 0.05$~mag (5 per cent in luminosity).

In accordance with ``Arnett's Rule''  
(Arnett 1982; Arnett, Branch \& Wheeler 1985) -- which states that the emitted
bolometric luminosity is roughly equal to the instantaneous rate of
generation of radioactive
luminosity at maximum light -- the increase in peak luminosity
resulting from the 3D structure is accompanied by a faster rise to
maximum; the peak luminosity occurs approximately one day earlier in
the 3D calculations than the 1D case.

The 3D model was also used to look for viewing-angle dependency of
the light curve. Light curves were extracted for a set of eight
randomly chosen viewing angles and compared to the angle-averaged
light curve. No significant ($> 1 - 2$ per cent) departures from the average
light curve were found; this can be understood since large scale
departures from sphericity are not present in this model,
partially as a result of the assumed reflection symmetries (see above). 

It is interesting to note that the results from the model considered
here show that inhomogeneity would act in
the sense of pushing model results closer to observation; 
for example, a recent comparison of light curves computed from 1D models 
with observations (Blinnikov et al. 2006) 
does suggest that the 1D models produce light curves which rise too
slowly and underestimate peak luminosities (see their figures 8 -- 13).
Direct
comparisons with the calculations of Blinnikov et al. (2006) are not
possible here owing to the considerably more complex treatment of
opacity that they
adopt, but it may reasonably be speculated that their rise times would
also be shorter if multi-dimensional effects were incorporated.

\begin{figure}
\epsfig{file=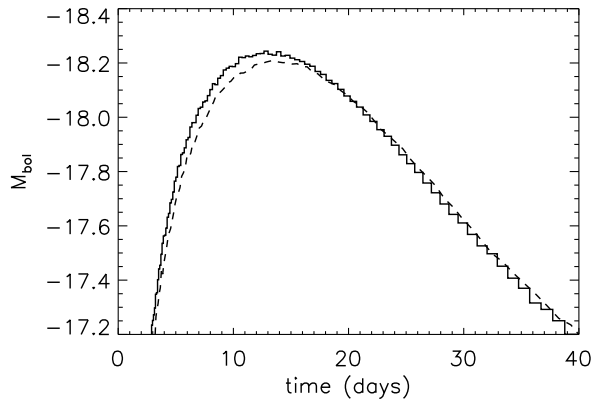, height=6cm}
\caption{
The light curve obtained from the 3D inhomogeneous
model described in Section~4.1 (solid histogram) and the comparison
spherically averaged model (dotted line). The light curve computed
from the 3D model does not vary significantly with viewing angle.
}
\end{figure}

The scale of the effect here is rather too small to directly have major
implications for the confrontation of models and observations. However,
the effect is not completely negligible and it is plausible that the
calculations presented here may underestimate the true scale of this
phenomenon; in particular, both the reduction in resolution from that used
in the explosion model and the grossly simplified one-parameter treatment of
the opacity may both suppress 3D density and composition effects that
would be present in a complete treatment. Furthermore, only one explosion
model has been considered here and this model is constructed from a
simulation describing only one octant (see above). 
In the future,
it will be necessary to examine a
range of models including fully 3D models where the
greater complexity may enhance the effects.

\section{Discussion}

We have described and tested a new 3D, time-dependent Monte Carlo code for
modelling radiation transport in SNIa. The code adopts the methods
presented by Lucy (2005) and also incorporates a scheme which uses
Monte Carlo radiation field estimators to allow observables to be
extracted for specific viewing angles; this approach helps to suppress 
Monte Carlo noise and eliminates the need to obtain intensities by
angular binning of emergent Monte Carlo packets. 

This code has been used to investigate two classes of three
dimensional effects in SNIa models. First, two elliptical SNIa
toy models were used to investigate how large scale asphericity might
alter observable light curves. As expected from previous simplified
treatments (e.g. H\"{o}flich 1991), it was found that light
curves were brighter when viewed down the minor-axis than the
major-axis. The brightness enhancement is largest at early times and
becomes smaller during the decline phase as the supernova ejecta
becomes less optically thick.
For a model with axis ratio comparable to that suggested
by polarisation data for real SNIa (Howell et al. 2001; Wang et al. 2003), 
the differences in peak brightness
and light curve shapes between viewing angles is detectable and in
principle could play a role in the observed scatter about the mean
relationship between light curve brightness and width.

Secondly, a model with structure based on the results of recent
three-dimensional hydrodynamical simulations was used to study the
effect of inhomogeneities in both density 
and composition on light curves. It was found that 
3D structure could lead to light curves which are both
brighter at early times and which peaked sooner after the explosion
that is found from 1D models. For the particular model considered,
the effect was rather modest ($\approx$ 5 per cent around maximum
light) but study of other explosion models is required to quantify 
the possible diversity in this effect.

Considerable further work is required to fully understand the role of
3D effects on radiation transport in SNIa. 
The use of the grey approximation in the treatment of {\sc uvoir} 
radiation is currently the greatest limiting factor to the practical
applicability of these results. 
In calculations involving a uniform 
grey absorption coefficient, the line-of-sight opacity depends 
solely on the total column density.
Under realistic conditions for SNIa ejecta, the opacity is dominated
by spectral lines (see e.g. Pinto \& Eastman 2000b) 
and is thus a strong function of frequency, velocity
gradient, composition and the ionization/excitation
state of the plasma. Indeed, when strong lines dominate, the opacity
has little direct sensitivity to the column density and is instead
mostly a function of the density of spectral lines in frequency
space. Furthermore, photon escape from highly opaque material is
facilitated by frequency redistribution to those regions of the
spectrum where there are relatively few spectral lines 
(see e.g. Pinto \& Eastman 2000b for a detailed discussion); in the
grey calculations this means of escape is not available and all the
energy packets must burrow through the imposed opacity. Given this, it
is likely that the grey approximation overestimates the role of
geometry in determining photon propagation. However, further
calculations are required to determine and understand this in
detail. In the context of the inhomogeneous model discussion in
Section~4, a more realistic treatment of the role played by
compositional inhomogeneity would be of particular interest; only
a crude attempt at describing the variation of opacity with
composition has been used here (equation~8) and a sophisticated treatment
involving the very real differences between the atomic properties of
different elements may lead to interesting effects.

Fortunately, as discussed by Lucy (2005), the Monte Carlo method is
well suited to incorporating realistic physics in a complex geometry
which will allow the interplay of more detailed micro-physics and 3D
structure to be studied in the near future;
this would
involve a more sophisticated treatment of the opacity, perhaps similar
to the methods employed by Blinnikov et al. (2006) or Kasen et
al. (2006). 
From such a calculations it would be possible to extract not only a
bolometric light curve, but also light curves for specific photometric
bands, an important step in the progression towards confrontation of
theory and observation.

Light curve calculations also need to be performed for a
wider range of explosion models than the exploratory set discussed
here. This is necessary in order to quantify how the various effects
might differ in scale depending on various properties of the explosion
and so help to understand what role they might play in establishing
both the average properties and diversity of SN observations.

Eventually, it would also be valuable to undertake a
similar assessment of the role of 3D effects in full NLTE modelling of 
{\sc uvoir}-spectra; much of the greatest diagnostic power lies in
spectral modelling -- including both intensity and polarimetry -- and
thus it is important that commonly used modelling assumptions, such as that of
spherical symmetry, are studied in the context of modern explosion
models.

\section*{Acknowledgements}
 
I am very grateful to F. R\"{o}pke for making available the data used
for the trial 3D model described in Section~4 and for advice
relating to hydrodynamical models and their
interpretation; to W. Hillebrandt for useful comments and suggestions
relating to various aspects of this work; to L. Lucy for making
available the numerical results of his test light curve computations
(used for Figure~1); and to D. Sauer and R. Kotak for helpful discussions.
I also thank the referee, P. Pinto, for his careful reading of this
manuscript and constructive comments and suggestions.

\section*{References}

Arnett W. D., 1982, ApJ, 253, 785\\
Arnett W. D., Branch D., Wheeler J. C., 1985, Nature, 314,\\
\indent 337\\
Ambwani K., Sutherland P., 1988, ApJ, 325, 820\\
Benetti S., et al., 2004, MNRAS, 348, 261\\
Benetti S., et al., 2005, ApJ, 623, 1011\\
Blinnikov S. I., R\"{o}pke F. K., Sorokina E. I., Gieseler M.,\\
\indent Reinecke M., Travaglio C., Hillebrandt W., Stritzinger \\
\indent	M., 2006, A\&A, 453, 229\\
Branch D., Falk S. W., Uomoto A. K., Wills B. J., McCall\\
\indent  M. L., Rybski P., 1981, ApJ, 244, 780\\
Branch D, 1992, ApJ, 392, 35\\
Gamezo V. N., Khokhlov A. M., Oran E. S., Chtchelkanova \\
\indent A. Y., Rosenberg R. O., 2003, Science, 299, 77\\
H\"{o}flich P., 1991, A\&A, 246, 481\\
H\"{o}flich P., Wheeler J. C., Thielemann F. K., 1998, ApJ,\\
\indent 495, 617\\
Howell D. A., H\"{o}flich P., Wang L., Wheeler J. C., 2001,\\
\indent  ApJ, 556, 302\\
Hoyle F., Fowler W. A., 1960, ApJ, 132, 565\\
Kasen D., Thomas R. C., Nugent P., 2006, ApJ, 651, 366\\
Kozma C., Fransson C., Hillebrandt W., Travaglio C., \\
\indent Sollerman J., Reinecke M., R\"{o}pke F. K., Spyromilio J.,\\
\indent 2005, A\&A, 437, 983\\
Lucy L. B., 1999, A\&A, 345, 211\\
Lucy L. B., 1987, in ESO Workshop on the SN 1987A, ed.\\
\indent I. J. Danziger, 417, European Southern Observatory,\\
\indent Garching\\
Lucy L. B., 2005, A\&A, 429, 19\\
Maeda K., Mazzali P. A., Nomoto K., 2006, ApJ, 645, 1331\\
Mazzali P. A., Lucy L. B., 1993, A\&A, 279, 447\\ 
Mazzali P. A., Podsiadlowski Ph., 2006, MNRAS, 369, L19\\
Niemeyer J. C., Hillebrandt W., Woosley S. E., 1996, ApJ,\\
\indent 471, 903\\
Nomoto K., Thielemann F.-K., Yokoi K., 1984, ApJ, 286,\\
\indent 644\\
M\"{u}ller E., Arnett W. D., 1986, ApJ, 307, 619\\
Phillips M. M., 1993, ApJ, 413, L105\\
Pinto P. A., Eastman R. G., 2000a, ApJ, 530, 744\\
Pinto P. A., Eastman R. G., 2000b, ApJ, 530, 757\\
Reinecke M., Hillebrandt W., Niemeyer J. C., 2002, A\&A,\\
\indent 391, 1167\\
R\"{o}pke F. K., 2005, A\&A, 432, 969\\
R\"{o}pke F. K., Gieseler M., Reinecke M., Travaglio C.,\\
\indent Hillebrandt W., 2006, A\&A, 453, 203\\
Wang L., et al. 2003, ApJ, 591, 1110\\
Yoon S.-C., Langer N., 2005, A\&A, 435, 967\\

\label{lastpage}

\end{document}